      [1999/12/01 v1.4c Il Nuovo Cimento]
\documentclass{hcimento}

\input{hepsfig.sty}

\title{GRB afterglows: deep Newtonian phase and its application}
\author{Y.F.~Huang\from{ins:nju}\ETC,
T.~Lu\from{ins:purple},
        \atque
K.S. Cheng\from{ins:hku}\thanks{Corresponding author: Y.F. Huang (hyf@nju.edu.cn).}}
\instlist{\inst{ins:nju} Department of Astronomy, Nanjing University, Nanjing 210093, China
  \inst{ins:purple} Purple Mountain Observatory, Chinese Academy of Sciences, Nanjing 210008, China
  \inst{ins:hku} Department of Physics, the University of Hong Kong, Hong Kong, China
}
\PACSes{
\PACSit{95.30.Qd}{Magnetohydrodynamics and plasmas.}
\PACSit{97.60.Bw}{Supernovae}
\PACSit{97.60.Lf}{Black holes}}

\begin{document}

\maketitle

\begin{abstract}
Gamma-ray burst afterglows have been observed for months 
or even years in a few cases. It deserves noting that at such late stages, the 
remnants should have entered the deep Newtonian phase, during 
which the majority of shock-accelerated electrons will no longer 
be highly relativistic. To calculate the afterglows, we must assume that the 
electrons obey a power-law distribution according to their kinetic 
energy, not simply the Lorentz factor. 
\end{abstract}

\section{Introduction}
Gamma-ray bursts (GRBs) have been recognized as the most relativistic phenomena 
in the Universe. In 1997, Wijers et al. once 
discussed GRB afterglows of the non-relativistic phase \cite{WijersR}. However, for quite 
a long period, many authors were obviously beclouded by the energetics 
of GRBs and emission in the non-relativistic phase was generally omitted. 
In 1998, Huang et al. stressed the importance of the Newtonian phase for
the first time \cite{HuangD98}. In fact, the Lorentz factor of GRB blastwave evolves as 
$   \gamma \approx (200 - 400) E_{51}^{1/8} n_0^{-1/8} t_{\rm s}^{-3/8}$ 
in the ultra-relativistic phase. It is clear that the shock will no longer be 
ultra-relativistic within tens of days. Today, the importance of non-relativistic 
phase has been realized by more and more 
authors \cite{LivioW,FrailW,DermerB,DermerH,Piro,intZand,PanaitescuK,ZhangM,Tiengo04,Frail05}. 
For example, in the famous case of GRB 030329, the transition to the non-relativistic
regime is believed to be detected, since its X-ray and radio afterglow light
curves flattened achromatically at $t \sim 40$ --- 50 day \cite{Tiengo04,Frail05}.

\begin{figure}[htb]
  \begin{center}
  \leavevmode
  \centerline{ 
  \epsfig{figure=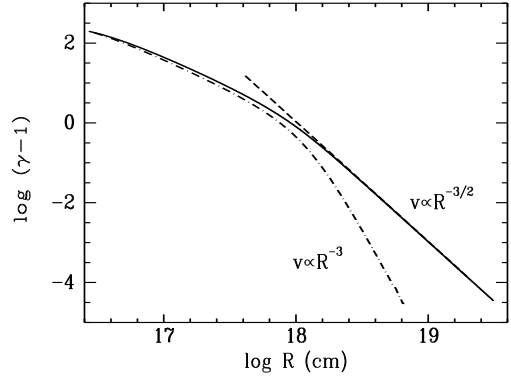,width=3.0in,height=2.0in,angle=270,
  bbllx=125pt, bblly=170pt, bburx=525pt, bbury=720pt}
  }
\caption {Evolution of the bulk Lorentz factor ($\gamma$) vs. radius for an isotropic fireball
    \cite{Huang}. The dashed line illustrates the famous Sedov solution for a
    Newtonian fireball.  The solid line is plot according to Huang et al.'s 
    generical dynamical model, which is correct in both the relativistic and 
 the Newtonian phases \cite{HuangD99}.  The dash-dotted line shows the result of an older dynamical model, 
    which is not correct in the Newtonian phase. }
  \end{center}
  \end{figure}

Recently it was further noted that GRB afterglows may enter the deep Newtonion phase 
typically in a few months \cite{HuangC}, when the minimum Lorentz factor of 
shock-accelerated electrons ($\gamma_{\rm e,min} \sim \xi_{\rm e} (\gamma-1) m_{\rm p}/m_{\rm e}$)
becomes less than a few. At this stage, most electrons will cease to be 
ultra-relativistic and their distribution function needs to be reconsidered \cite{HuangC}. 

\section{Model}
To describe the deceleration of GRB ejecta,  we use the refined generic 
dynamical model proposed by Huang et al. \cite{HuangD99}, which
is mainly characterized by
\begin{equation}
\label{eq2}
\frac{d \gamma}{d m} = - \frac{\gamma^2 - 1}
       {M_{\rm ej} + \epsilon m + 2 ( 1 - \epsilon) \gamma m}.
\end{equation}
Fig. 1 shows clearly that this equation is applicable in both the 
ultra-relativistic phase and the Newtonian phase. Detailed description 
concerning the overall dynamical evolution of isotropic fireballs as well as 
collimated jets can be found in Huang et al. \cite{HuangD99,HuangG,HuangD00}.

Shock-accelerated electrons are usually assumed to distribute as 
$d N_{\rm e}'/d \gamma_{\rm e} \propto \gamma_{\rm e}^{-p}$.
However, in the deep Newtonian phase, most electrons are 
non-relativistic. 
To calculate afterglows, the distribution 
function now needs to be revised as \cite{HuangC}
\begin{equation}
\label{eq10}
{d N_{\rm e}'}/{d \gamma_{\rm e}} \propto (\gamma_{\rm e} - 1)^{-p}.
\end{equation}
Optical afterglows can then be calculated conveniently by integrating 
synchrotron emission from those electrons with Lorentz factors above 
a critical value ($\gamma_{\rm e,syn}$) \cite{HuangC}. 

\begin{figure}[htb]
  \begin{center}
  \leavevmode
  \centerline{ 
  \epsfig{figure=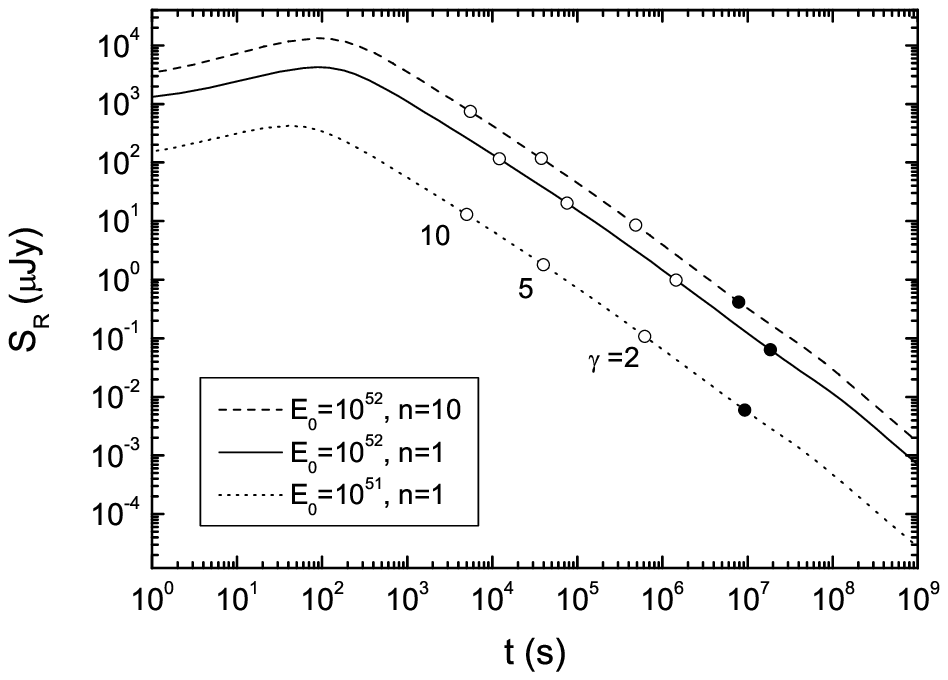,width=3.0in,height=1.85in,angle=0,
  bbllx=-40pt, bblly=20pt, bburx=290pt, bbury=215pt}
  \hspace{0.2in}
  \epsfig{figure=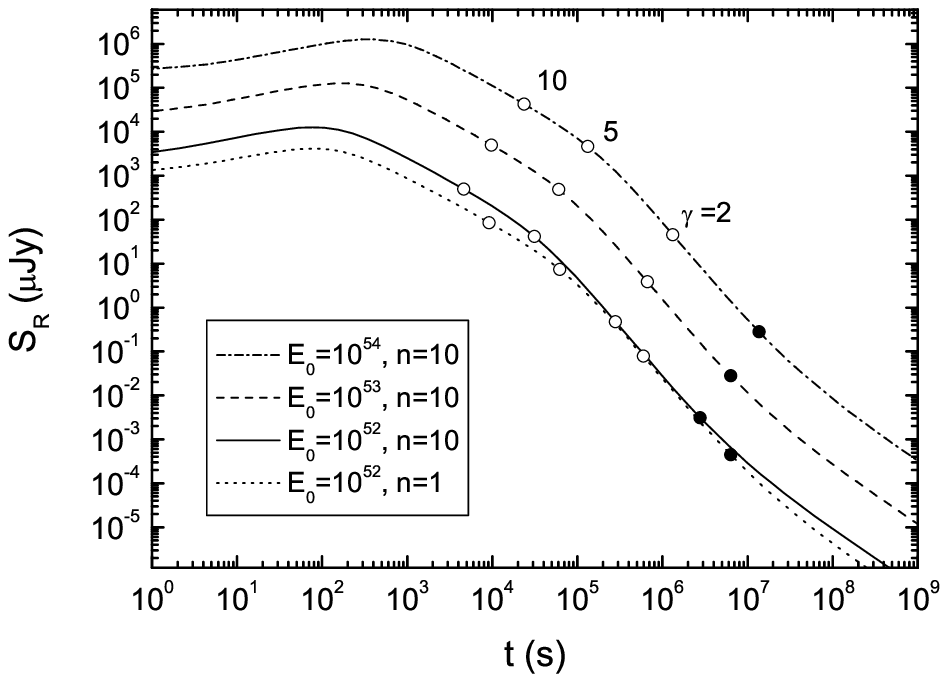,width=3.0in,height=1.85in,angle=0,
  bbllx=15pt, bblly=20pt, bburx=345pt, bbury=215pt}
  }
\caption {R-band optical afterglows from isotropic fireballs 
({\bf a}, left panel) and conical jets ({\bf b}, right panel) \cite{HuangC}. 
The black dot on each light curve indicates the moment when 
$\gamma_{\rm e,min} = \gamma_{\rm e,syn} \equiv 5$, and 
open circles mark the time when the bulk Lorentz 
factor $\gamma = 2,$ 5 and 10 respectively.}
  \end{center}
  \end{figure}

Detailed numerical results for isotropic fireballs and highly collimated 
conical or cylindrical jets have been presented by Huang \& Cheng \cite{HuangC}. 
Here we show some exemplar results in Fig 2. Fig 2a illustrates optical afterglows
from isotropic fireballs. Note that the deep Newtonian phase typically begins at 
about $10^7$ s. The light curves steepen slightly after that.
Our results are consistent with analytical solutions.
On the contrary, afterglow light curves of conical jets (Fig. 2b) flatten in 
the deep Newtonian phase, which is also consistent with analytical 
solutions.

\section{Application}

It has been shown clearly that GRB ejecta enters the deep Newtonian phase 
typically in about 3 months. Afterglows in the deep Newtonian phase are thus
very important, especially in the following three cases. Case 1, late
afterglows. Optical afterglows from some GRBs have been detected for more than six
months. Radio afterglows are detectable even three years
later. The deep Newtonian phase is unavoidable when such observations are to 
be accounted for. Case 2, GRBs with a dense medium. For some GRBs, the density of circum-burst medium
may be as large as $10^3$ cm$^{-3}$, or even $10^6$ cm$^{-3}$ in some rare
cases. The GRB ejecta will then decelerate very rapidly and may enter the deep 
Newtonian phase in less than 20 days. Case 3, fireballs with relatively small 
initial Lorentz factor. This includes failed GRBs and the two-component jet model
of GRBs, which will be discussed in more details below. 

Failed GRBs \cite{HuangD02}, or dirty fireballs as named by Dermer et al. \cite{Dermer99}, 
are relativistic fireballs with initial Lorentz factor $\gamma_0 \ll 100$ --- 1000. 
They cannot produce normal GRBs, but may give birth to X-ray flashes and contribute 
to orphan afterglows. The simple discovery of orphan afterglows then does not 
necessarily mean that GRBs are highly collimated \cite{HuangD02}, although this 
was once regarded as a hopeful method for measuring the beaming angle
of GRBs \cite{Rhoads,GranotP}. 
To judge whether an orphan afterglow comes from a failed GRB or a jetted but 
off-axis GRB, Huang et al. suggested that the most important thing
is to monitor the orphan for a relatively long period \cite{HuangD02}. Obviously, 
the calculation of afterglows in the deep Newtonian phase is necessary in such 
studies. 

\begin{figure}[htb]
  \begin{center}
  \leavevmode
  \centerline{ 
  \epsfig{figure=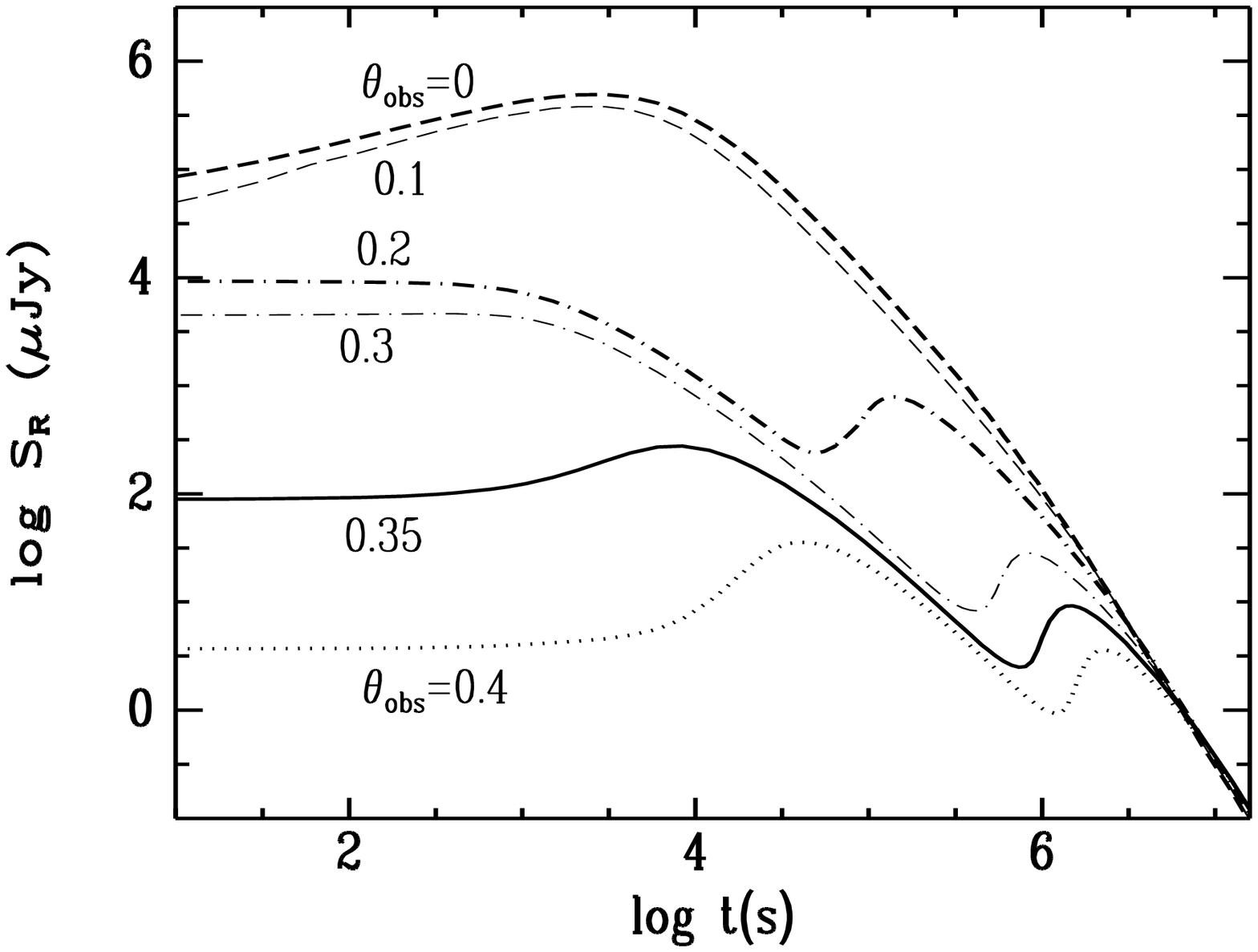,width=3.0in,height=1.85in,angle=270,
  bbllx=130pt, bblly=10pt, bburx=510pt, bbury=620pt}
  \hspace{0.2in}
  \epsfig{figure=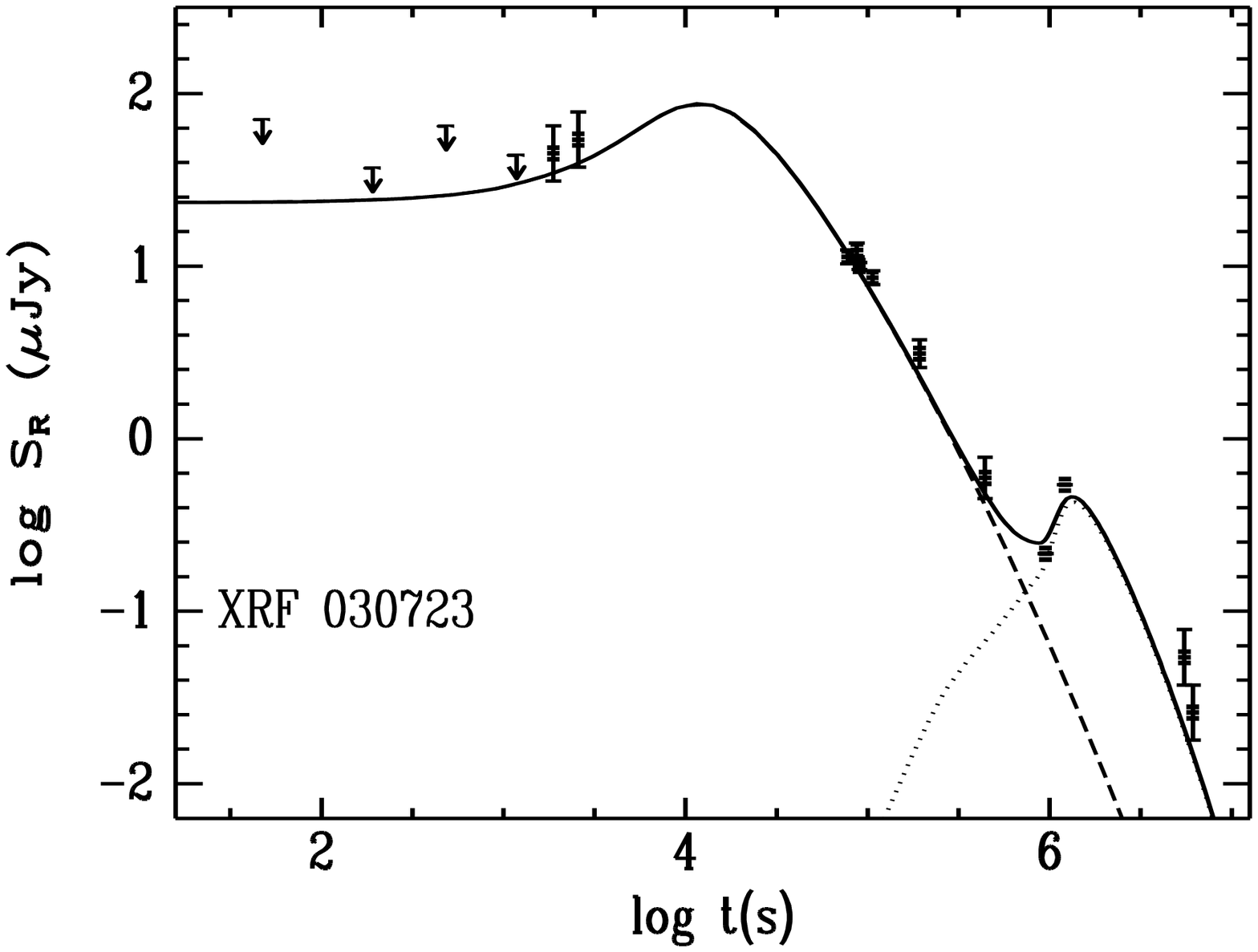,width=3.0in,height=1.85in,angle=270,
  bbllx=130pt, bblly=120pt, bburx=510pt, bbury=730pt}
  }
\caption {Effects of the viewing angle on the afterglow from a two-component jet 
({\bf a}, left panel), and its fit to the optical afterglow of XRF 030723 
({\bf b}, right panel) \cite{Huang04}.}
  \end{center}
  \end{figure}

Recently it was proposed that some GRB jets may have two components: a central narrow 
ultra-relativistic outflow and an outer, wider, mildly relativistic ejecta 
\cite{Berger03,Huang04}. This two-component jet model can potentially give a unified 
description for GRBs and X-ray flashes \cite{Huang04}: if our 
line of sight is within the narrow component, a normal GRB will be observed; On the 
contrary, if the line of sight is outside the narrow component but within the wide 
component, an X-ray flash will be witnessed. In both cases, long-lasting afterglows
can be detected. In such a model, since the outer ejecta is midly relativistic 
at the beginning, radiation in the deep Newtonian phase will 
be inevitably involved in calculating its afterglows.  
Afterglow behaviors of two-component jets have been studied 
detailedly by Huang et al. \cite{Huang04}. Here, as an example, we illustrates 
the effects of the viewing angle on the optical light curves in 
Fig 3a. Fig 3b shows clearly that the two-component jet model can give a 
perfect explanation for the observed optical afterglows from the X-ray flash XRF 030723.

\acknowledgments
This research was supported by the National Natural Science Foundation
of China (10003001, 10221001, 10233010 and 10473023), 
the FANEDD (Project No: 200125), the National 973 
Project, and a RGC grant of Hong Kong SAR.

\end{document}